# Discrete Quantum Mechanics I

## Quantum Covariance

Charles Francis

**Abstract:**
   Classical general covariance is found from the idea that a vector is a physical quantity which exists independently of choice of coordinate system and is unchanged by a change of coordinate system. It is often assumed that there exists some form of absolute mathematical space or space-time, and that in a flat space approximation vectors can be imagined between defined points in this space-time, much like arrowed lines drawn on paper. However, while classical vector quantities can be represented on paper, in the quantum domain physical quantities do not in general exist with precise values except in measurement; a change of apparatus, for example by rotating it, may affect the outcome of the measurement, so the condition for general covariance need not apply. The purpose of this paper is to re-examine covariance in the context of an orthodox (Dirac-Von Neumann) interpretation of quantum mechanics and to find a new formulation, here called quantum covariance, to express the fundamental principle that the laws of physics are everywhere the same, to show that the quantum covariant formulation of quantum mechanics is the most precise expression of the orthodox interpretation, and to show that this formulation allows the unification of general relativity with quantum mechanics for non-interacting particles.

Charles Francis MA (Cantab) PhD (Lond)
Lluest, Neuaddlwyd
Lampeter
Ceredigion  UK
SA48 7RG

**Affiliation:** Jesus College, Cambridge

## 1   Background

The question as to whether measured distances are a prior property of empty space, or whether they are simply relationships found in matter has been open since the introduction of absolute space in Newton's Principia and the criticisms levelled against it by Leibniz and others. Although the mathematical formulation of physical law has depended on an assumption of space, or more recently spacetime, imbued with mathematical properties, the Leibnizian relationist view (which had previously been advocated by Descartes and has its roots in Aristotle, Democritus and other Greek writers) continues to hold intellectual appeal and there is some reason both within foundations of quantum mechanics and in relativity for thinking that this would be the correct way to formulate physical theory [7]. In recent years the Leibnizian view has been advocated by Smolin [25], Rovelli [23] and others as motivation for work on background free theories such as spin networks and spin foams. This paper seeks to show that the relationist view has profound implications for the definition of reference frames, necessitating a new formulation of covariance in the quantum domain, and that, when quantum covariance is correctly applied, quantum theory and general relativity can be unified in a single model giving the predictions of both to current limits of experimental accuracy.

**Unification**

The axioms or fundamental postulates of this model will be presented in the form of empirical definitions and definitional truisms. Empirical definitions define quantities observed in nature and contain the physical assumptions of the model, while definitional truisms determine mathematical structure. We are not here concerned with the philosophical distinction and both will be labelled simply as definitions. Unification requires the correct set of postulates containing the essential features of relativity and quantum mechanics and giving the predictions of both while omitting (or modifying) assumptions which lead to the widely recognised incompatibilities between them. This paper uses a subset of the standard assumptions of quantum mechanics and relativity, replacing the background space $\mathbb{R}^n$ which appears in standard formulations of quantum mechanics with a relativistic analysis of measurement using the *k*-calculus, and replacing the assumption of a Schrödinger equation, Lagrangian, or Hamiltonian with a determination of time evolution from covariance. There will be minimal discussion of omitted assumptions and the reader must be responsible for recognising that the mathematical model discussed here follows from the given definitions and is not necessarily compatible with ideas contained in other accounts. However the reader may observe that in each case an empirical assumption based directly on observation is preferred to a metaphysical one which cannot be directly observed but can only be induced from the correctness of physical prediction. There is twofold justification for the assumptions and definitions presented here, first from heuristic physical argument rooted in observation and second because its rigorous mathematical predictions agree with quantum mechanics in flat space and with classical general relativity in the respective approximations.

**Outline**

This paper starts in section 2, *Discrete Quantum Theory* with a reformulation of quantum theory in accordance with an orthodox (in the sense of Bub [4]) interpretation, that states in quantum mechanics describe our information about a system rather than the physical system itself. For clarity it will be called information space interpretation. Information is derived from measurement and the interpretation is that quantum theory is essentially a theory of probabilistic relationships between measurement results not a model of physical processes between measurements. This interpretation follows Dirac [8] and Von Neumann [26]. It has its origins in the Copenhagen interpretation as discussed by Heisenberg [14], and has much in common with modern views such as Mermin [16], Adami and Cerf [1], and Youssef [28] (although here probability theory is exotic only in the sense that it is an extension of classical probability, and does not require Bayesian interpretation). As in Copenhagen interpretation matter has an unknown but real behaviour which is not directly described by quantum mechanics. The state describes not what is, but our knowledge of what might happen in measurement, by giving a probability for each outcome. The laws of vector space represent a weighted logical OR between potential outcomes of hypothetical measurement in many valued logic while the inner product is a means of calculating probabilities for statements relating initial information and final measurement. Natural units will be used, $c = 1$, $\hbar = 1$ ($G \neq 1$).

Section 3, *Relativity* reviews the *k*-calculus. There are accounts of the *k*-calculus in [2] and [3] and a brief account is given in [6]. It is reviewed here because Bondi's *k*-calculus had a context of classical e.m. radiation and was used to introduce only special relativity. It will be regarded here in a more abstract context, with arguments based ultimately on a maximum theoretical speed of information rather than the speed of light, and adapted to a quantum treatment of general relativity. As used by Bondi, *k* is Doppler red shift (*z* will not be used and $k = 1 + z$ will be called red shift). Although this is not an elementary paper this will be an elementary account to ensure that fundamental definitions and concepts are precisely expressed as is required for a the description of quantum covariance. Curved space is introduced together with macroscopic parallel displacement, used in quantum covariance.

It has long been known that there is no finite unitary representation of the Lorentz group but section 4, *The Principle of Quantum Covariance* argues that as information is observer dependent it is necessary to reinterpret covariance. The relationship of quantum covariance to manifest covariance is discussed, and any empirical difference is too small to detect by direct measurement. In the quantum covariant formulation wave functions will be defined as equivalence classes on an infinite family of finite dimensional Hilbert spaces, with time evolution determined by the requirement for covariance. It is shown that geodesic motion follows for classical motions.

Section 5, *Gravity* defines inertial motion and discusses the principle of equivalence then extends the *k*-calculus to include gravitational red shift and to account for gravity by observing that, after introducing a minimum time delay in the transfer of information, the metric defined by the radar method leads to Schwarzschild geometry for a single elementary charged particle and obeys Einstein's field equation generally in the classical limit.

## 2 Discrete Quantum Theory

**Reference Matter**

When a human observer seeks to quantify nature he chooses some particular matter from which to define a reference frame or chooses certain matter from which he builds his experimental apparatus. He then observes a defined relationship between this specially, but arbitrarily, chosen reference matter and whatever matter is the subject of study. Measurement is here distinguished from a simple count of a number of objects, and is defined to mean a count of units of a measured quantity, where the definition of the unit of measurement invokes direct comparison between some aspect of the subject of measurement and a property of the reference matter used to define the unit of measurement. The division between reference matter and subject matter is present in all measurement and appears as the distinction between particle and apparatus in quantum mechanics, and in the definition of position relative to a reference frame in special relativity. Reference matter is not necessarily inertial. It may define a reference frame or it may take some other form, such as the apparatus used for an experiment in quantum mechanics. For example it may be a cyclotron, or a linear accelerator, or it may include a photographic plate. But whatever we measure in the universe we measure it relative to a property of other, arbitrarily chosen, matter in the universe, and what we actually determine in measurement is the relationship between the two, rather than an absolute property of the subject of measurement.

The specification of reference matter is to a large degree arbitrary and D'Inverno [6] defines a reference frame as a clock, a ruler and coordinate axes, whereas Rindler [22] describes reference frame as a "conventional standard" and discusses the attachment of a frame to definite matter, such as the Earth or the "fixed" stars while Misner, Thorne and Wheeler [17] define proper reference frame as a Minkowski coordinate system with a given clock at the origin. Whatever reference matter is used there must be some form of clock and some means of specifying position. A clock may be anything from an atomic clock to one's own subjective sense of time, and a ruler may be a 'rigid' rod (together with the means to move the rod into position), it may be an entire radar system, or it may simply be a subjective judgement of distance based on a parallax measurement carried out by one's eyes.

It is not necessary to treat the apparatus from a classical perspective, as in the standard Copenhagen interpretation. We merely note that the result of a measurement of position at given time is always three numbers, and we use those three numbers to label a condition found in matter. In this view geometry is simply and literally world (*geo-*) measurement (*-metry*), as it was for the Greeks; it follows that to understand geometry we must study the nature and process of measurement, not assume a pre-existent manifold with geometrical properties. Thus space-time is described not as a background into which matter is placed in the manner of Newtonian absolute space, but as an observer dependent set of potential measurement results. As Rovelli [23] says, "*According to Descartes, there is no 'empty space'. There are only objects and it makes sense to say that an object A is contiguous to object B. The 'location' of an object A is the set of the objects to which A is contiguous. 'Motion' is change in location*". This is in strict accordance with the orthodox interpretation of quantum mechanics. In Dirac's words "*In the general case we cannot speak of an observable having a value for a particular state, but we can.... speak of the probability of its having a specified value for the state, meaning the probability of this specified value being obtained when one makes a measurement of the observable*" [8]. When this statement is applied to the position observable, it follows that precise position only exists in measurement of position, and hence that there is no ontological background geometrical space or space-time.

**Coordinates**

We are particularly interested in the measurement of time and position which is sufficient for the study of many (it has been said all) other physical quantities, and we restrict our treatment to those physical quantities which can be reduced to a set of measurements of position. For example a classical measurement of velocity may be reduced to a time trial over a measured distance, and a typical measurement of momentum of a particle involves plotting its path in a bubble chamber, being a set of positions over a time interval. Any apparatus has a finite resolution. The values written down are *n*- tuples of terminating decimals which can be scaled to integers in units of the resolution of the apparatus. Margins of error can be represented as defining either a finite set of such integers, or as the limit of the resolution of the apparatus itself. In practice there is also a bound on the magnitude of each coordinate. Without loss of generality the same bound, $\nu \in \mathbb{N}$, is used for each coordinate. Knowledge of the state is thus restricted to this set of *n*-tuples and the results of measurement of position are in a (subset of a) finite region $N \subset \mathbb{N}^3$. in units of the resolution of the apparatus. We use a synchronous coordinate system, bounded both by size and by the resolution of measuring apparatus, with time as a parameter as in non-relativistic quantum mechanics. Without loss of generality define

**Definition:** *The coordinate system is* $N = (-\nu, \nu] \otimes (-\nu, \nu] \otimes (-\nu, \nu] \subset \mathbb{N}^3$ *for some* $\nu \in \mathbb{N}$, *where* $(-\nu, \nu] = \{x \in \mathbb{N} | -\nu < x \leq \nu\}$.

This account is in essence a description of particles. It is sometimes assumed that a particle always has a definite, if unknown, position but this is not the case here, since a value for the position observable is not assumed to exist between measurements. We have

**Definition:** *A particle is any physical entity whose position can be measured such that the result of such measurement is a value* $x \in N$. *An elementary particle is one which cannot physically be subdivided into particles for which separate positions can be measured even in principle.*

There is no significance in the bound, $\nu$, of a given coordinate system $N \subset \mathbb{N}^3$. Matter outside of N is ignored. If matter goes outside of N it is merely moving out of a coordinate system, not out of the universe and (in the absence of singularities) it is possible to describe its motion in another coordinate system with another origin. Thus each coordinate system is defined as a lattice determined by the finite resolution of the apparatus; it is bounded by practical considerations and is part of information space of a particular observer, not prior ontology. N is not assumed rectilinear. Any lattice can be used as appropriate to the measuring apparatus of an individual observer. Not every element of N need correspond to a possible measurement result, but N contains the set of possible measurement results for a measurement of position with a given apparatus. Even if it is not intended to take the limit $\nu \to \infty$, we assume that N may chosen large enough to say with certainty up to the limit of experimental accuracy, that it contains any particle under study for the duration of the experiment. More strictly data is discarded from any trial in which there is not both a well defined initial and final state, and the probability amplitudes defined below refer to conditional probabilities such that both initial and final states are unambiguously determined. Thus conservation of probability is imposed as usual (and incidentally there is no detection loophole in Bell tests [13] - in the absence of unambiguous detection this model does not apply). Wave functions will be defined not on a particular lattice but on the set of possible lattices, and we define the family

**Definition:** *Let $\mathcal{N}$ be the set of finite discrete coordinate systems of sufficient size to include any particles under study and which can be defined by a given observer by means of physical measurement.*

**Information Space**

**Definition:** *For any $x \in N$, $|x\rangle$ is the ket corresponding to a measurement of position with result x at time t. $|x\rangle$ is called a position ket.*

In the absence of information we cannot describe the actual configuration of particles; kets are names or labels for the information we have about a state, not the physical state itself. The significance is that the principle of superposition will be introduced as a definitional truism in a naming system, not as a physical assumption. Although kets are not states of matter, but merely names for states, we loosely refer to kets as states in keeping with common practice when no ambiguity arises.

In a typical measurement in quantum mechanics we study a particle in near isolation. The suggestion is that there are too few ontological relationships to generate the property of position. Then position does not exist prior to the measurement, and the measurement itself is responsible for introducing interactions which generate position. In this case, prior to the measurement, the state of the system is not labelled by a position ket, and we define labels containing information about other states – specifically the information about what would happen in a measurement.

**Definition:** *A vector space, $\mathbb{H} = \mathbb{H}(t)$, is constructed over $\mathbb{C}$ with basis $\{|x\rangle | x \in \mathbb{N}\}$. The members of $\mathbb{H}$ are called kets.*

**Definition:** *$\forall |f\rangle, |g\rangle \in \mathbb{H}$, the braket $\langle g|f \rangle$ is the hermitian form defined by its action on the basis:*

$$\forall x, y \in \mathbb{N}, \langle x|y \rangle = \delta_{xy} \qquad 2.1$$

**Theorem:** It is routine to prove that

$$\langle g|f \rangle = \sum_{x \in \mathbb{N}} \langle g|x \rangle \langle x|f \rangle, \text{ and hence formally } \sum_{x \in \mathbb{N}} |x\rangle\langle x| = 1 \qquad 2.2$$

**Definition:** *The position function of the ket $|f\rangle \in \mathbb{H}$ is the mapping $\mathbb{N} \to \mathbb{C}$ defined by*

$$\forall x \in \mathbb{N}, x \to \langle x|f \rangle$$

Later the position function will be identified with the restriction of the wave function to $\mathbb{N}$, but we use the term position function, because it is discrete and because a wave motion is not assumed.

**Probability Interpretation**

In scientific measurement we set up many repetitions of a system described by the initial state, and record the frequency of each result, or final state. Probability is interpreted here simply as a prediction of a frequency distribution, so a mathematical model must associate a probability with each possible result. The probability of a given result can be used to attach a label to the state by observing that

$$P(x|f) = \frac{|\langle f|x\rangle|^2}{\langle f|f \rangle} \qquad 2.3$$

has the properties of a probability distribution, and by associating the ket $|f\rangle \in \mathbb{H}$ with any state such that $\forall x \in \mathbb{N}$ the probability that a measurement of position has result $x$ is given by $P(x|f)$. Probabilities defined from 2.3 are invariant under local $U(1)$ gauge symmetry and interactions must preserve this symmetry since it describes redundant information appearing in the mathematical model but without physical meaning.

**Quantum Logic**

The position function (and more generally the scalar product) can be identified with a proposition in a many valued logic [21]. The value of the position function is the (complex) truth value of the proposition. Classical logic applies to sets of statements about the real world which are definitely true or definitely false. For example, when we make a statement

$$\mathscr{P}(x) = \text{\textit{The position of a particle is x}} \qquad 2.4$$

we tend to assume that it is definitely true or definitely false. Such statements are said to be sharp, meaning that they have truth values from the set $\{0, 1\}$. If it is actually the case that $\mathscr{P}(x)$ is definitely either true or false then classical logic and classical mechanics apply. But in quantum mechanics it is not, in general, possible to say that a particle has a position when position is not measured, and in the absence of actual measurement we can only consider sets of propositions describing hypothetical measurement results, such as the set of propositions of the form:

$$\mathscr{T}(x) = \text{If a measurement of position were to be done the result would be } x \qquad 2.5$$

The only empirical states are those which are measured, and only for measured states can we make sharp statements with in the form of $\mathscr{P}(x)$. But we can categorise other states according to the likelihood of what might happen in measurement by using the structure of Hilbert space and a probability interpretation based on the inner product. Thus we construct new propositions by identifying the vector sum $a|f\rangle + b|g\rangle$ with weighted logical OR between propositions and we say that $\langle x|f\rangle$ is the truth value of the proposition $\mathscr{T}(x)$. Multiplication by a scalar only has meaning as a weighting between alternatives and we have that $\forall |f\rangle \in \mathbb{H}$, $\forall \lambda \in \mathbb{C}$ such that $\lambda \neq 0$, $\lambda|f\rangle$ labels the same state (or set of states) as $|f\rangle$. So long as we recognise that Hilbert space is just a set of rules, a calculational device which says nothing about metaphysics there is no inconsistency, ambiguity or other problem with the use of kets to label unmeasured states in this way, the property of superposition coming from the logic not the metaphysic.

We can make statements in the form of $\mathscr{T}(x)$ and interpret the probability amplitude for such statements as a truth value, but we cannot in general make statements in the form $\mathscr{P}(x)$. This begs the question *what is wrong with* $\mathscr{P}(x)$? Is there no such thing as numerical position except in certain states, or is there no such thing as a particle? We will see that if, in fact, there is no background continuum then there is no reason to reject the notion that the universe is composed of particles, but we do reject the notion that the position of a particle is meaningful except when it is defined through relationships created through interaction with other particles, including (but not exclusive to) relationships determined by a measurement process. Before continuing with the formulation of quantum mechanics we consider the implications of procedures for the measurement of position, bearing in mind that a requirement of any such procedure is that it can be calibrated to give the same result as some defining procedure, and that we are at liberty to choose the defining procedure.

**Collapse**

Without loss of generality let $|f\rangle$ and $|g\rangle$ be normalised $\langle f|f\rangle = \langle g|g\rangle = 1$. Generally the probability that a measurement will give the result $f$ given a previous measurement result $g$ is

$$P(g|f) = |\langle f|g\rangle|^2 \qquad 2.6$$

provided also that there is no other information about the state (which would alter the probability). New information always causes a change of state, not because of a physical collapse, but because the state is simply a label for the available information. The collapse of the wave function is simply the the modification of a probability to a conditional probability when a condition becomes known. The information space interpretation then inverts the measurement problem: collapse represents a change in information due to a new measurement but Schrödinger's equation requires explanation – interference patterns are real. The requirement for a wave equation will be found in section 4, *The Principle of Quantum Covariance*.

In a measurement of position, the ket describing the initial state $|f\rangle$ of the apparatus is changed into a ket describing a position in $X$, a region of space of size determined by the measuring apparatus. The operator effecting the change is

$$Z(X) = \sum_{x \in X} |x\rangle\langle x| \qquad 2.7$$

as is shown by direct application

$$Z(X)|f\rangle = \sum_{x \in X} |x\rangle\langle x|f\rangle \qquad 2.8$$

since the resulting state is a weighted logical or between positions in $X$. Applying $Z$ a second time to 2.8 and using 2.2 shows $Z(X)$, is a projection operator $Z(X)Z(X) = Z(X)$ reflecting the observation that a second measurement of a quantity gives the same result as the first. Then

$$\langle f|Z(X)Z(X)|f\rangle = \sum_{x \in X} \langle f|x\rangle\langle x|f\rangle \qquad 2.9$$

Then by 2.3, 2.9 is the sum of the probabilities that the particle is found at each individual position, $x \in X$. In other words it is the probability that a measurement of position finds the particle in N. In the case that $X$ contains only the point $x$, $X = \{x\}$, 2.8 becomes

$$Z(x)|f\rangle = |x\rangle\langle x|f\rangle \qquad 2.10$$

Thus the position function, $\langle x|f\rangle$, can be interpreted as the magnitude of the projection from the state $|f\rangle$ of the apparatus into the state $|x\rangle$, or the component of $|f\rangle$ on the basis ket $|x\rangle$.

**Uncertainty**

Classical probability theory describes situations in which every parameter exists, but some are not known. Probabilistic results come from different values taken by unknown parameters. We have a similar situation here. In this model there are no relationships between particles apart from those generated by physical interaction and we do not know the precise configuration of particle interactions. An experiment can be described by a large configuration of particles incorporating the measuring apparatus as well as the process being measured. The configuration of particles has been partly determined by setting up the experimental apparatus, giving a set of possible outcomes to a measurement but the precise pattern of interactions which leads to each particular measurement result is unknown. It is impossible to determine every detail of the configuration since the determination of each detail requires measurement, which in turn requires a larger apparatus containing new unknowns in the configuration of particles. Thus there is a residual level of uncertainty, which can never be removed by experiment. The interpretation of 2.8 is that the probability that the interactions combine to $Z(x)$, is

$$\langle f|Z(x)Z(x)|f\rangle = \langle f|Z(x)|f\rangle = \langle f|x\rangle\langle x|f\rangle = |\langle x|f\rangle|^2 \qquad 2.11$$

Then 2.6 can be understood as a classical probability function, where the random variable runs over the set of projection operators,

$$Z(x) = |x\rangle\langle x| \qquad 2.12$$

Then each $Z(x)$ represents a set of unknown configurations of particle interactions in measurement, namely that set of configurations which lead to the result $x$.

**Observable Quantities**

The above argument generalises immediately to the values found in measurement of other quantities. Now each eigenstate represents the set of unknown configurations of particle interactions for which the corresponding eigenvalue is the result of the measurement. Performing a measurement requires the configuration to be in a state for which the measurement has a definite result and is equivalent to the application of a projection operator. Under the identification of addition with quantum logical OR the expectation of all the results is a hermitian operator equal to a weighted sum over a family of projection operators.

**Definition:** *For any real valued function or functional $O(x)$ the observable operator or observable, O, is defined by*

$$\forall |f\rangle \in \mathbb{H} \quad O|f\rangle = \sum_{x \in N} |x\rangle O(x) \langle x|f\rangle \qquad 2.13$$

*Formally*

$$O = \sum_{x \in N} |x\rangle O(x) \langle x| \qquad 2.14$$

Clearly observables are hermitian and a standard treatment of eigenvalues and eigenstates can be given. It is not assumed that practical measurement is possible for any observable $O$. The information from an experiment for which well defined value can be read as the result (such as the measurement of momentum in a bubble chamber) collapses the state to an eigenstate of the corresponding observable operator.

**Theorem:** The expectation of an observable $O$ in the state $|f\rangle \in \mathbb{H}$ is

$$\langle O \rangle = \sum_{x \in N} \langle f|x\rangle O(x) \langle x|f\rangle = \langle f|O|f\rangle \qquad 2.15$$

*Proof:* Left as an exercise. Write $O$ as the sum of projection operators onto its eigenstates.

**Classical Correspondence**

In classical physics we study the behaviour of large numbers of elementary particles, calculated statistically using 2.15 such that a classical property is the expectation of the corresponding observable. For example the centre of gravity of a macroscopic body is a weighted average of the positions of elementary particles, and, as will be shown in *Discrete Quantum Mechanics II*, Maxwell's equations apply to the expectations of the photon field operator and the Dirac current operator. Classical physics is found as the limit of large sample behaviour so that the effect of observation is negligible and time evolution of a classical property is equal to that of the expectation of an observable operator. The classical correspondence is the large sample limit $\hbar/N \to 0$ as sample size $N \to \infty$ (not the limit $\hbar \to 0$ as is sometimes stated; Planck's constant is simply a change of scale from natural to conventional units and it would be meaningless to let it go to zero).

**Effective Measurement**

In this model the existence of a value for an observable quantity depends only on the configuration of matter. If the interactions in a configuration of matter combine to a process describable by an eigenstate of an observable operator then the value of the corresponding observable quantity exists independently of observation or measurement, and is given by the corresponding eigenvalue. However the state is our information about the system and is an eigenstate only if we know the value of the quantity. Schrödinger's cat is definitely either alive or dead because it is a classical cat and we can apply to it classical laws statistically derived from 2.15, but the state is a superposition until the box is opened. Here quantum mechanics does not describe the evolution in time of a physical wave function, but describes the probability relationship between an initial measurement result at time $t_1$ and a final measurement result at time $t_2$. In classical physics there is sufficient information (from 2.15) to determine the motion at each instant between the initial and final state (up to experimental accuracy). Since intermediate states are determinate and may be calculated in principal they may effectively be regarded as measured states. A classical motion may effectively be described as a sequence of measured states at instances separated by some time interval $\chi$ which is sufficiently small that there is negligible alteration in predictions in the limit in which $\chi$ tends to zero. The state at each instant may be regarded as the initial state, and state at the next instant may be regarded as the final state, which becomes the initial state for the next part of the motion.

## 3 Relativity

**The *k*-Calculus**

Hermann Bondi expressed the view that '*the Michelson-Morley experiment is a mere tautology"* [3], and Einstein is also said to have thought the result of this experiment to be inevitable [20], the reason being that reference frames require light for their definition and coordinates are thus calibrated to the speed of light, not the other way around. In fact this is not quite true, in part because the photon could have (or acquire) a non-zero but immeasurably small mass, but principally because in quantum field theory the amplitude for the creation of a particle and its annihilation at any non-synchronous point is non-zero, even outside the light cone, so the speed of individual photons is not constrained. Nonetheless we can discuss the maximum theoretical speed of information, which does not depend on the properties of any one particle, and it is almost inevitable that the maximum speed of information is an absolute constant from the following argument. If we are to measure the time and distance of an event spacially separated from ourselves, then information must travel between us and the event. If we know the speed of information transfer, we can determine the time and distance of the event. But speed is defined in terms of time and distance, which leads to a paradox. The distance of an event must be known before we can talk of speed, but the speed of information must be known to determine the time and distance of the event. To resolve the paradox we must find something fundamental and base everything else on it. The maximum speed of information is either bounded or unbounded. If it is unbounded we must allow instantaneous action at a distance and laws of physics which are in conflict with those we observe. If we reject that then we may conclude that there is a maximum speed of information, which we can call *c*. We must use a speed of information transfer to synchronise separated clocks because even if we were to synchronise them by bringing them together, we have no other guarantee that they will remain synchronised once separated. Thus, although the Michelson-Morley experiment is not tautology, it is tautologous to say that the maximum speed of information is the same (up to scaling) in all reference frames, because there is no reference frame which does not depend for its definition on the maximum speed of information. In practice light does travel at *c*, to the limits of experimental accuracy, and for the purpose of this paper it will be taken that light is the carrier of information and also travels at *c*.

**Idealisation**

In the *k*-calculus the radar method is used to determine time and distance coordinates, as described below under *Reference Frames*. This will be taken as the definition of space-time coordinates since any other method of measurement can be calibrated to give an identical result. Radar is preferred to a ruler, because it applies directly to both large and small distances, and because a single measurement can be used for both time and space coordinates. The *k*-calculus actually uses an idealisation of radar. It is imagined that a radar pulse can be sent in any given direction at a precise time and that a reflected signal returns after a interval which can be precisely timed. There is no such thing as a perfectly confined wave packet, but, as previously remarked, the definition of time and space coordinates depends on the maximum theoretical speed of information in any direction, not on practical issues of signalling with e.m. radiation. For example the simplest possible antenna, the dipole antenna, has a transmission/receiving pattern that resembles a figure 8. The transmitting pattern introduces an uncertainty in the direction in which photons are transmitted, and there is a corresponding uncertainty in our ability to determine the direction from which a received photon came. The smaller the object we are trying to detect by radar, the more its radiation pattern resembles that of a dipole. Hence there is a relationship between the uncertainty in the position of an object and the size of the object compared to the wavelength of the probing signal. Indeed a radar pulse can be thought of as a wave packet describing the uncertainty in time of transmission of a photon. To reduce uncertainty to achieve perfect eigenstates of position would require radar signals of infinitesimal wavelength and infinite energy. It is legitimate to discuss such idealisations since the definition of a metric does not depend on practical issues but on a bounding value, the maximum theoretical speed of information, which can only be attained in the theoretical limit of a radar pulse of infinitesimal duration. It is reasonable to apply the analysis to light in the limiting case of eigenstates of position and to use a space-time diagrams showing the reflection of a photon as sharply defined (in the sense of many valued logic) because the analysis for a general wave function can be recovered from the analysis of eigenstates by superposition.

**Reference Frames**

There is room for confusion between two very similar questions, 'What is time?' and 'What is the time?'. The first question has something to do with consciousness, and our perception of time as a flow from past to future. It admits no easy answer, but is quite distinct from the second question and only the second question is relevant in the definition of space-time coordinates. The answer to the question 'What is the time?' is always something like 4:30 or 6:25.

**Definition:** *The time in physics is a number given by a clock.*

There are many different types of clock, but every clock has two common elements, a repeating process and a counter. The rest of the mechanism converts the number of repetitions to conventional units of time and allows synchronisation with other clocks. A good clock should provide accurate measurement and it should give a uniform measure of time. We cannot count less than one repetition of the process in the clock, so for good resolution the process must repeat as rapidly as possible.

**Definition:** *In a uniform clock, the repeating process must repeat each time identical to the last, uninfluenced by external matter.*

Until 1967 the uniformity of the measurement of time was assured by the rotation of the earth around the sun. Because of the masses of celestial objects compared with the energy of matter and radi-ation falling on them it was possible to neglect perturbations to the motion and use the absence of perturbation to define uniform time. However the basic unit of the year had to be subdivided by means of other clocks. Since 1967 the standard second has been by definition 9,192,631,770 periods of the radiation from the unperturbed microwave transition between the two hyperfine levels of the ground state of $Cs^{133}$. There is clearly a logical problem with a discussion of atomic clocks before a full development of quantum mechanics, but it is logically possible to develop the model from the pre-1967 definition, and we may comment pre-emptively on the new definition.

Actually a Caesium clock does not count periods of radiation, the output is a count of oscillations of a quartz crystal locked at 5MHz by a series of feedback loops using electronic multipliers to produce microwave radiation at a frequency to maximise the energy transitions between the hyperfine levels. The repeating process is a period of the oscillation of the quartz together with all such feedback loops, and, despite the figure given in the definition of the second, the output of the clock cannot be resolved to greater accuracy than 5MHz. The maximisation of the energy transitions is merely the means by which we observe that the process has repeated identically, under the assumption that the laws of physics remain the same, and in particular that Planck's constant remains the same in determining the relationship between frequency of radiation and energy (c.f. the discussion in [17] to the effect that uniform time is that which makes the laws of physics simple).

The definition of the standard second raises a significant conceptual issue. 9,192,631,770 periods of radiation are actually periods of an unobservable phase ($U(1)$ gauge symmetry) which appears only in information space and cannot be directly attributed to a physical oscillation. The measurement is one of energy, not time, and is related to time via Planck's constant. The most basic form of energy is the bare mass of an elementary particle, which is also a constant of nature, so it raises the question as to whether a more fundamental definition of time could (or should) be taken directly from the bare mass of a particle. This question will not be fully addressed here, but the indication in this paper is that proper time is a fundamental physical property of each particle, but that it is related to rate of interaction and has an inverse relationship to non-physical phase.

**Definition:** *Clocks may be called identical if they measure identical units of time when they are together, and if moving them does not noticeably affect the physical processes of their operation.*

In particular Caesium clocks use an identical physical process wherever they are so the standard second is identical everywhere when it is measured by a clock at the position of the measurement.

A clock defines the time, but does so only at one place. A space-time reference frame also requires a definition of distance, and a definition of time at a distance from the clock. Both are provided for by the radar method.

**Definition:** *The distance of an event is half the time for light to go from a clock to the event, to be reflected and to return to the clock. The time of reflection is the mean of time sent and time of return.*

**Definition:** *A reference frame is the set of possible results of time, distance and direction measurements based on a given clock together with whatever other reference matter is required to carry out measurement of time and position.*

It does not need to be specified whether coordinates are polar or cartesian or other, since we may change coordinate system with a matrix transformation, and for convenience a reference frame is described as the set of coordinate systems defined from particular reference matter including a clock.

**Definition:** *In Minkowski coordinates a reference frame has one time and three cartesian space axes.*

**Inertial Matter**

If we are to use a particular reference frame as the fundamental object for a description of matter and describe physics in terms of its properties we must ensure that the properties of the reference frame are entirely due to the reference matter used for its definition and that there may be no affect on physical law as measured in the frame due to external influences. This may be done in practice by physically isolating the matter, or empirically by observing external influences to be negligible. We define:

**Definition:** *An inertial object is one whose motion is not affected by direct action of other matter (to the limits of experimental accuracy).*

**Definition:** *An inertial reference frame is one whose origin and coordinate axis are defined to be constant with respect to inertial matter (in particular the clock is inertial).*

These definitions should be noted carefully, since one sometimes finds confusion in the literature about the definition of inertia. In some accounts 'inertial' is defined from Newton's first law to mean uniform motion. Such a definition is not used here. The literal latin meaning of inertia is "no action" and this is the meaning which will be adopted in this paper. Newton's first law will be demonstrated as a local approximation applicable to inertial reference frames from the general principle of the homogeneity of physical law, as in Noether's theorem. Here Newton's first law is a physical law about the behaviour of inertial matter, not the definition of it.

**Space-time Diagrams**

In the information space interpretation motion takes on Descartes' meaning, change of location, and contiguous becomes synonymous with interaction. Each particle follows a precise but unknown path, where the path strictly consists of a series of interactions with other particles, not a set of points in space-time. The position of a particle (to the extent to which it has one) is a figment of interactions with other particles defining the relationship between the particle and reference matter. In the case of an eigenstate of definite position measured by the radar method, a reflected photon follows a definite path as shown in figure 1, the probability amplitude for other paths being zero. The physical path is the set {emission, reflection, absorption}. The rest of the diagram is pure mathematical construction; the diagram is defined such that lines of equal time are horizontal and lines of equal distance are vertical and light is drawn at 45°.

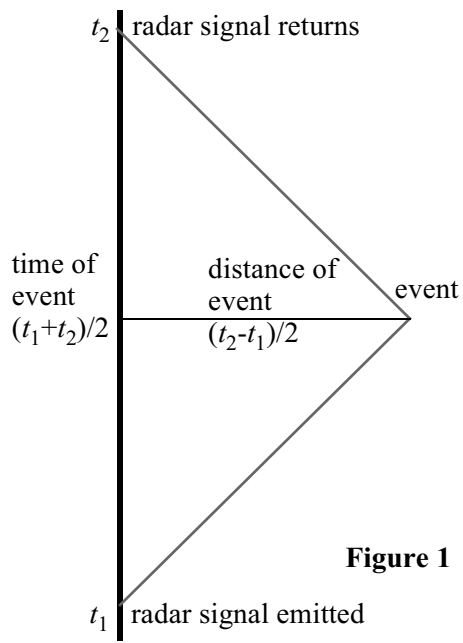

**Figure 1**

Isotropy means local spherical symmetry about any origin. The radar method measures direction and, since for inertial coordinates there is no preferred direction, space-time diagrams provide a 3 dimensional treatment in spite of their apparently one dimensional nature. Space-time diagrams may be understood as showing the radial coordinate in 3 dimensions (it is trivial also to see that in even cartesian coordinates the algebra of the radial coordinate of a space-time diagrams is formally the same as that of vectors in any number of dimensions).

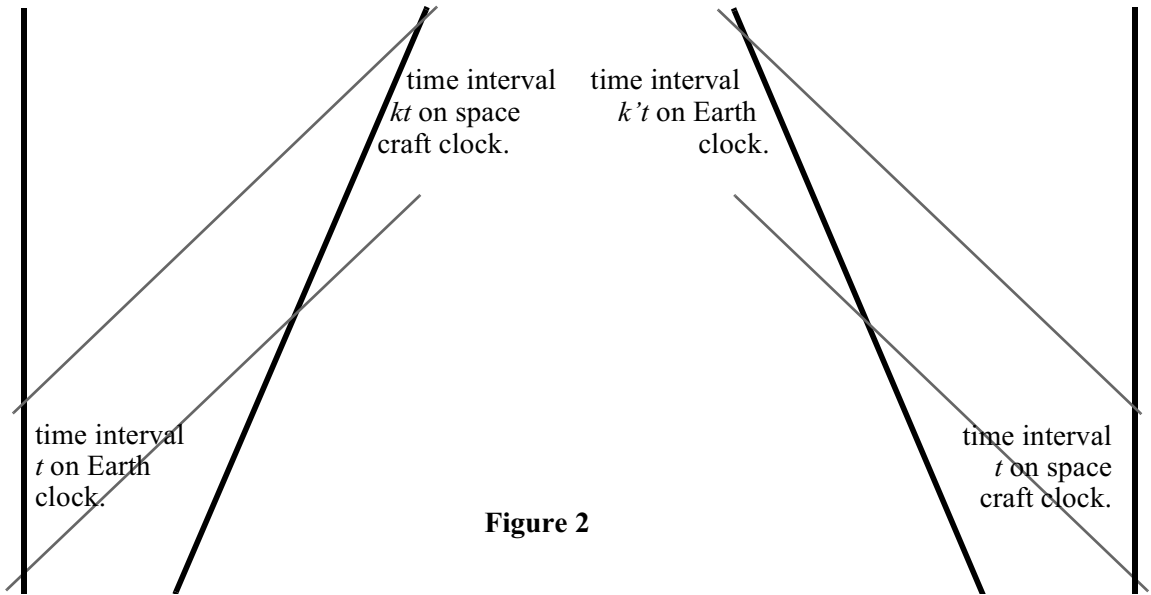

**Figure 2**

If we wish to compare our coordinate system with the coordinate system of a moving observer, we need to know what unit of time the moving observer is using. Once clocks are separated, there is no way to synchronise them directly, but, by the definitions given above, two clocks will give an identical unit of time if the physical processes in each are the same. Figure 2 shows a space craft moving uniformly in the Earth's reference frame. The space craft and Earth have identical clocks and communicate with each other by radio or light. The Earth sends the space craft two signals at an interval $t$. The space craft receives them at an interval $kt$ on the space craft's clock for some $k \in \mathbb{R}$. By considering the signals as the start and stop of a burst of light of a set number of wavelengths of a set frequency $k$ is immediately recognisable as the red shift factor. Similarly if the observer on the space craft sends two signals at an interval $t$ on his clock, they are received at an interval $k't$ on the Earth.

Provided only that they are inertial there is no fundamental difference between the matter in the space craft and the matter in the Earth. The space craft can be regarded as stationary, and the Earth as moving. Red shift for signals sent by Earth and received by the space craft is not generally the same as that sent by the space craft to the earth since there may be effects due to location with respect to other matter. But if there is no action on a body it has no preferred orientation in space-time and for inertial frames $k = k'$ whenever their origins coincide. The defining condition for the special theory of relativity is that we use reference frames such that

**Definition:** *In Minkowski coordinates red shift between inertial frames is both constant and equal for both observers, $k = k'$.*

From isotropy we expect that Minkowski space-time applies as a local approximation everywhere and for the remainder of this section it is assumed that $k = k'$, this being the condition for the special, rather than the general, theory of relativity. The treatment here is restricted to showing the formula for time dilation and length contraction, since it is widely known (and straightforward to show) that the properties of space-time vectors and Lorentz transformation can be derived directly from them.

**Theorem:** (Time dilation figure 3) The time $T$ measured by a space craft's clock during an interval $t$ on the Earths clock is given by

$$T = t\sqrt{1-v^2} \qquad 3.1$$

*Proof:* The space craft and the Earth set both clocks to zero at the moment the space craft passes the Earth. The space craft is moving at speed $v$, so by definition, after time $t$ on the Earth clock, the space craft has travelled distance $vt$. Therefore Earth's signal was sent at time $t - vt$, and returned at time $t + vt$. For inertial reference frames, if the space craft sends the Earth signals at an interval $t$ the Earth receives them at an interval $kt$. So

$$T = k(t - vt). \qquad 3.2$$

Then by applying the Doppler shift again for the signal coming back

$$t + vt = k^2(t - vt) \qquad 3.3$$

Eliminating $k$ gives 3.1, the formula for time dilation.

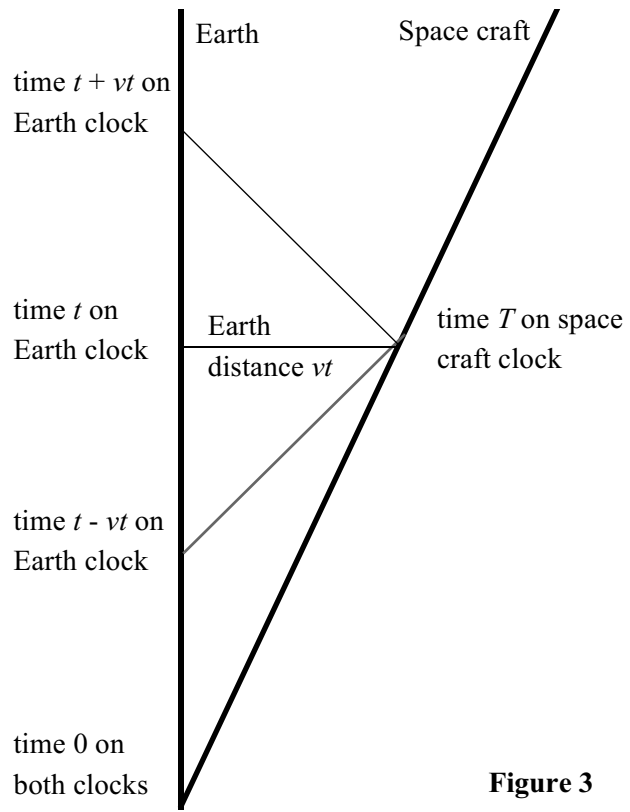

Figure 3

**Theorem:** (Lorentz Contraction, figure 4) A distance $d$ on the earth is measured on a space craft to be

$$D = \frac{d}{\sqrt{1-v^2}} \qquad 3.4$$

*Proof:* The bow and stern of the space craft are shown as parallel lines. The space craft's clock is in the bow. The space craft and Earth set their clocks to zero when the bow passes the Earth clock. Earth uses radar to measure the distance, $d$, to the stern, by sending a signal at time $-d$, which returns at time $d$ on the Earth clock. The same signal is used to measure $D$ on the spaceship. The outgoing signal passes the bow at time $-(d/k)$ on the space craft's clock, and the returning signal reaches the bow at time $kd$. So, according to the moving space craft

$$D = (kd + d/k)/2 \qquad 3.5$$

Eliminating $k$ using 3.3 gives 3.4.

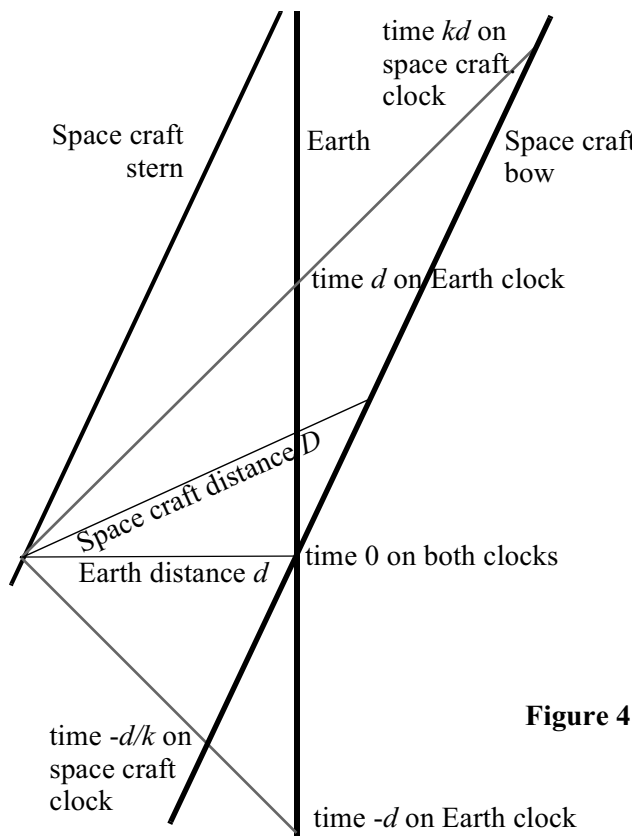

Figure 4

**Non-Euclidean Geometry**

As has been said isotropy implies that in the limit of coincident origins the red shift measurement of one inertial frame on another is equal to that of the second frame on the first, so that space-time is locally Minkowski. General relativity considers larger regions such that the distance between the origins of local reference frames cannot be neglected and the red shift measurements are not equal. Under such conditions the flat space formulae of special relativity following from 3.1 and 3.4 break down and the following treatment develops the notion of the curvature of space-time in terms of the stretching or distortion of maps (or charts) drawn in coordinate space.

**Coordinate Space Vectors**

Let coordinate space be denoted by axes $\alpha$ for $\alpha = 0, 1, 2, 3$ with Minkowski metric, $\eta_{\alpha\beta}$ everywhere. $\eta_{\alpha\beta}$ is not the physical metric, but is an abstract metric used for mapping. Up to changes of scale $\eta_{\alpha\beta}$ corresponds to the Euclidean metric of, for example, the paper in a Mercator projection, as distinct from the spherical metric of geography. Since coordinate space is flat it is identical to itself under changes of scale and can be used to draw maps of regions of space-time, analogous to the drawing of maps of the surface of the Earth on flat paper. Then curvature is naturally conceived in terms of the scaling distortions in the map. In general straight lines in coordinate space are not geometrically straight, but for a sufficiently short line segment the deviation from straightness is not detectable, so to first order a short rod placed at $x$ will appear as a small displacement vector $A^\alpha$ defined as usual as the difference in the coordinates of one end of the rod from the other. A coordinate space vector is defined by inverting the scaling distortions of the map so that the coefficients of a coordinate space vector are equal to the coefficients of the corresponding vector in primed local Minkowski coordinates: To do this we first define the matrix

$$k^\mu_{\ \nu} = x^{\mu'}_{,\nu} \qquad 3.6$$

which transforms the unprimed axes to the local primed Minkowski coordinates, then we write

**Definition:** *For the vector $A^\alpha$ at position x the corresponding coordinate space vector, barred to distinguish it from an ordinary, or physical, vector, is defined by*

$$\bar{A}^\alpha(x) = k^\alpha_{\ \beta}(x) A^\beta(x) \qquad 3.7$$

A position coordinate can be regarded as a coordinate space vector, being the displacement vector from the origin in coordinate space, but it does not usefully correspond to a physical vector and the bar can be omitted for position vectors $\bar{x} = x$. Because the coefficients of a coordinate space vector are equal to its coordinates in local Minkowski space-time 3.7 can be rewritten in primed local Minkowski coordinates at $x$

$$A^{\alpha'}(x) = x^{\alpha'}_{,\beta}(x) A^\beta(x) \qquad 3.8$$

This is a tensor equation, so the coordinate space vector is independent of the choice of coordinate space. The scalar product of the vectors $A^\alpha$ and $B^\alpha$ is defined at $x$ as usual by

$$A \cdot B = g_{\alpha\beta}(x) A^\alpha B^\beta \qquad 3.9$$

For coordinate space vectors it implicit that the metric is η, so the scalar product between the corresponding coordinate space vectors $\bar{A}^\alpha$ and $\bar{B}^\alpha$ is given by

$$\bar{A} \cdot \bar{B} = \eta_{\alpha\beta}\bar{A}^\alpha\bar{B}^\beta \qquad 3.10$$

Then

$$A \cdot B = g_{\alpha\beta}(x)A^\alpha B^\beta = \eta_{\mu\nu}k^\mu{}_\alpha(x)k^\nu{}_\beta(x)A^\alpha B^\beta = \eta_{\mu\nu}\bar{A}^\mu\bar{B}^\nu = \bar{A} \cdot \bar{B} \qquad 3.11$$

3.11 is true for any vectors $A^\alpha, B^\alpha$ so

$$g_{\alpha\beta}(x) = \eta_{\mu\nu}k^\mu{}_\alpha(x)k^\nu{}_\beta(x) \qquad 3.12$$

3.12 gives the metric in terms of the variable scale coefficients $k^\mu{}_\alpha(x)$ of coordinate space vectors compared to physical vectors.

**Observer Space-time**

An observer at O uses a locally Minkowski reference frame defined by axes α for α = 0, 1, 2, 3 (figure 5). The metric is Minkowski at O, and if the observer at the origin of a reference frame does not know, or does not wish to take account of, the metric at all points in space-time he may use radar to define simultaneity and Euclidean trigonometry to define a 'distance' between any two points, based only the measurements of distance and direction he can make where he stands. He will then have a space defined in a visible region of the universe with a constant Minkowski metric $\eta_{\alpha\beta}$ such that the physical metric $g_{\alpha\beta}(x)$ coincides with Minkowski metric at the origin. This will be called observer space-time at O. Observer space-time at $x$ is a coordinate space such that $g_{\alpha\beta}(x) = \eta_{\alpha\beta}$.

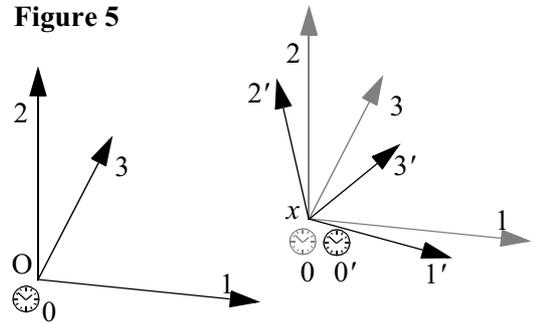

**Figure 5**

**Definition:** *Observer space-time consists of the possible results of measurement of space-time coordinates and calculations on them using Minkowski metric.*

Using Minkowski metric in observer space-time an observer does not find that a one metre rule is always one metre long, or that the laws of physics are everywhere the same (in fact bodies will accelerate at different rates depending on position). The 'real' metric is Minkowski at the origin and, in a small enough region bounded in time and space an observer cannot detect the difference between observer space-time and curved space-time at the limit of experimental accuracy. Observer space-time corresponds to the normal Newtonian way in which we most easily think of space and time, although in this model it is a part of information space, not prior ontology. It has the property of Newtonian absolute space that it is always and everywhere unchanging without relation to anything external, although it is not conceived as being "absolute" since there are different observer space-times at different origins.

Since the manifold in general relativity is locally Minkowski and observer space-time uses Minkowski metric it is natural to try and identify observer space-time and tangent space. In general this cannot be done in a rigorous manner because in Schwarzschild geometry there is singularity at any point-like gravitating particle and the tangent does not exist at a singularity. It is more true to say that observer space-time is the tangent to the manifold neglecting the gravitational effect of the matter used to define the reference frame, although in fact this statement is inverted because a manifold is not assumed as a prior, and is found from the collection of potential observer space-times.

**Parallel Displacement**

In standard treatments infinitesimal parallel displacements are used to develop the notion of parallel transport. We will also be interested in parallel displacement over finite distances. Unlike the more familiar parallel transport, parallel displacement does not invoke a notion of physically transporting a rod along a path, and is not path dependent. The justification for using parallel displacement rather than parallel transport in quantum relativity is twofold, first because of the known problems in field quantisation in curved space-time (see e.g. [27]) which may lead one to think that quantum fields cannot be consistently defined for a model based on particles, and second because, as will be seen, momentum is defined as a vector at the origin of an observer, A's, coordinates using the flat observer space metric; when the same vector is viewed in another observer, B's coordinates at a remote origin, it is mapped directly to a parallel vector at the origin of B's coordinates, not transported incrementally on a curved physical manifold from one observer to the other (this is essential to consistency if momentum is to be conserved since in general waves do not take a unique path from A to B).

A short rod placed at $x$ is described by vector $A(x)$. An identical short rod physically placed at $y$ so that its coordinate space vector is parallel to $\bar{A}(x)$. It is described by a vector $A(y)$ whose length is unchanged,

$$A^2(x) = A^2(y) \qquad 3.13$$

By parallel we mean that coordinate space components are proportional

$$k^\alpha{}_\beta(y)A^\beta(y) = \bar{A}^\alpha(y) \propto \bar{A}^\alpha(x) = k^\alpha{}_\beta(x)A^\beta(x) \qquad 3.14$$

In general this is not equality because of the different scaling distortions at $x$ and $y$ (although coordinate space is flat coordinate space vectors are defined at a point and cannot meaningfully be mapped into each other by translation). $A(y)$ is the result of parallel displacement of $A(x)$. If coordinate space is an observer space-time at $x$ we can without loss of generality choose coordinate axes to coincide with the primed axes of local Minkowski space at $x$, so 3.14 reduces to

$$k^\alpha{}_\beta(y)A^\beta(y) = \bar{A}^\alpha(y) \propto \bar{A}^\alpha(x) = A^\alpha(x) \qquad 3.15$$

Multiply both sides of 3.15 by $\eta_{\alpha\gamma}k^\gamma{}_\mu(y)$ and use 3.12

$$A_\mu(y) \propto A^\alpha(x)\eta_{\alpha\gamma}k^\gamma{}_\mu(y) = A_\gamma(x)k^\gamma{}_\mu(y) \qquad 3.16$$

Then the components of the right hand are equal to those of a vector

$$C_\mu = A_{\gamma'}(x)x^{\gamma'}_{,\mu}(y) \qquad 3.17$$

whose components in local primed Minkowski coordinates at $y$ are the same as those of $A_{\gamma'}(x)$. Thus the magnitudes of the left and right hand sides of 3.16 are the same and we have equality

$$A_\mu(y) = A_\gamma(x)k^\gamma{}_\mu(y) \qquad 3.18$$

This will be recognised as the standard formula for infinitesimal parallel displacement which is used to derive the formulae for parallel transport. We will also use it in the displacement from initial state to final state in quantum mechanics when this is not infinitesimal and it is not meaningful to intermediate displacements.

## 4   The Principle of Quantum Covariance

**General Covariance**

The principle of general relativity states *the laws of physics should be the same irrespective of the coordinate system which a particular observer uses to quantify them*. This is a form of the principle of homogeneity, that *the behaviour of matter is everywhere the same* (not to be confused with homogeneity of mass density as used in cosmology). Laws which are the same in all coordinate systems are most easily expressed in terms of invariants, quantities which are the same in all coordinate systems. The simplest invariant is an ordinary number or scalar. Another invariant is the tensor. A change of coordinates has no effect on a tensor, but it changes the description of a tensor in a coordinate system. The tensor transformation law (i.e. manifest covariance, or Lorentz covariance) may be found by defining displacement vectors in Minkowski space, and generalised to define general vector and tensor quantities which are the same in all coordinate systems in curved space-time. Then the form of the principle of homogeneity most directly applicable in relativity is the principle of general covariance, *The equations of physics have tensorial form.*

**Measurement of Vector Quantities**

Some form of the above argument for the principle of general covariance can be found in any book of general relativity. But the argument does not hold in the context of section 2, *Discrete Quantum Theory*. General covariance applies to classical vector quantities under the assumption that they are unchanged by measurement. But in quantum mechanics measured values arise from the action of the apparatus on the quantum system, creating an eigenstate of the corresponding observable operator. In discrete quantum theory a lattice describes the possible values taken from measurement by a particular apparatus, and a change of reference frame implies a change of apparatus (either by accelerating the apparatus or by switching to a different apparatus). The eigenstates of a vector observable in one reference frame are determined by the properties and resolution of a particular measuring apparatus, and cannot, in general, simultaneously be eigenstates of the corresponding vector observable in another frame using another apparatus (c.f. non-commutative geometry [5]).

**Quantum Covariance**

The broad meaning of *covariance* is that it refers to something which varies with something else, so as to preserve certain mathematical relations. If covariance is not now to be interpreted as manifest covariance or general covariance as applicable to the components of classical vectors, then a new form of covariance is required to express the principle of general relativity, that the laws of physics are the same in all reference frames. This will be called quantum covariance. Quantum covariance will mean that the laws of physics have the same form in any reference frame but does not mean that the same physical process may be described identically in different reference frames, since in quantum mechanics the reference frame, or the choice of apparatus, affects the process under study.

Measured values are interpreted as relationships between particle and apparatus so the laws of physics themselves are relationships between particle and apparatus, and they are expressed in terms of the properties of a given apparatus. In the discrete model the lattice is the representation of the apparatus and appears in the calculation of physical predictions only in the inner product, 2.2.

**Definition:** *Quantum covariance will mean the wave function is defined on a continuum, 4.21, while the inner product is discrete, 2.2, and that in a change of reference frames the lattice, N, and inner product appropriate to one reference frame are replaced with the lattice and inner product of another.* According to quantum covariance the sum in 2.2 is replaced in a change of reference frame, but we still require a means to transform the position function $\langle x|f \rangle$ as appropriate to the new frame. This will be given (in 4.23) after defining momentum space.

**Momentum Space**

Momentum will be defined mathematically from the properties of abstract vector space, not empirically. The correspondence with a measurable physical quantity is to be made using 3.7 and by demonstrating geodesic motion in the classical correspondence (section 5, *Gravity*). Although the formulae derived below are largely standard properties of the discrete Fourier transform they are reviewed here to define notations, to make clear that they are definitional properties of a mathematical structure not induced from empirical evidence, and to make clear that momentum is defined at the origin for each observer using the flat metric of observer space-time, not the physical metric of a space-time manifold.

**Definition:** *Momentum space is the 3-torus* $M = [-\pi, \pi] \otimes [-\pi, \pi] \otimes [-\pi, \pi] \subset \mathbb{R}^3$. *Elements of momentum space are called momenta.*

**Definition:** *For momentum* $p \in M$, *define a ket* $|p\rangle$, *by the position function*

$$\langle x|p \rangle = \left(\frac{1}{2\pi}\right)^{\frac{3}{2}} e^{ix \cdot p} \tag{4.1}$$

where the Euclidean metric of 3 dimensional observer space has been used. The expansion of $|p\rangle$ in the basis $\mathbb{H}_0$ is calculated by using the resolution of unity, 2.2

$$|p\rangle = \sum_{x \in N} |x\rangle\langle x|p\rangle = \left(\frac{1}{2\pi}\right)^{\frac{3}{2}} \sum_{x \in N} e^{ix \cdot p} |x\rangle \tag{4.2}$$

**Definition:** *For each ket* $|f\rangle$ *define the momentum space function as the Fourier transform*

$$F: M \to \mathbb{C} \text{ such that } F(p) = \langle p|f \rangle \tag{4.3}$$

By 2.2, $F$ can be expanded as a trigonometric polynomial

$$F(p) = \sum_{x \in N} \langle p|x\rangle\langle x|f\rangle = \left(\frac{1}{2\pi}\right)^{\frac{3}{2}} \sum_{x \in N} \langle x|f\rangle e^{-ix \cdot p} \tag{4.4}$$

by 4.1 and 4.3. By 4.4 and 4.1

$$\left(\frac{1}{2\pi}\right)^{\frac{3}{2}} \int_M d^3p\, F(p) e^{ix \cdot p} = \frac{1}{8\pi^3} \int_M d^3p \sum_{y \in N} \langle y|f\rangle e^{-iy \cdot p} e^{ix \cdot p} = \langle x|f\rangle \tag{4.5}$$

Rewriting 4.5 in the notations of 4.3 and 4.1

$$\langle x|f\rangle = \int_M d^3p\, \langle x|p\rangle\langle p|f\rangle \tag{4.6}$$

For any integrable $F':M \to \mathbb{C}$ there is a unique position function given by

$$\langle x|f\rangle = \left(\frac{1}{2\pi}\right)^{\frac{3}{2}} \int_M d^3p\, F'(p) e^{ix \cdot p} \qquad 4.7$$

4.7 is not invertible but defines an equivalence class of functions $F':M \to \mathbb{C}$ with the same position function.

**Definition:** *Members of this class are called representations of the momentum space wave function.*

Because the model is defined in position space with the hermitian product given by 4.5 all predictions are identical for each equivalent momentum space wave function. 4.4 picks out a unique invertible member of the equivalence class:

**Definition:** $F(p) = \langle p|f\rangle$ *is the analytic momentum space function in a given reference frame.*

By 2.2 and by 4.6 $\forall |f\rangle, |g\rangle \in \mathbb{H}$

$$\langle g|f\rangle = \sum_{x \in \mathbb{N}} \langle g|x\rangle\langle x|f\rangle = \int_M d^3p \sum_{x \in \mathbb{N}} \langle g|x\rangle\langle x|p\rangle\langle p|f\rangle = \int_M d^3p\, \langle g|p\rangle\langle p|f\rangle \qquad 4.8$$

4.8 is true for all $|f\rangle$ and $|g\rangle$, so we can define a second form of the resolution of unity

$$\int_M d^3p\, |p\rangle\langle p| = 1 \qquad 4.9$$

It follows immediately that

$$\langle q|f\rangle = \int_M d^3p\, \langle q|p\rangle\langle p|f\rangle$$

So $\langle p|q\rangle$ is a Dirac delta function on the test space of momentum space functions. Explicitly, by 4.2,

$$\delta:M \to \mathbb{C} \qquad \delta(p-q) = \langle q|p\rangle = \frac{1}{8\pi^3} \sum_{x \in \mathbb{N}} e^{ix \cdot (p-q)} \qquad 4.10$$

**Wave Mechanics**

In this treatment $\mathbb{H}$ is only a labelling system and its construction has required no physics beyond the knowledge that we can measure position coordinates and we can measure the relative frequency of each result of a repeated measurement. The description of physical processes in terms of this labelling system requires a law describing the time evolution of kets. This paper develops the non-interacting theory, which applies to any particle, fundamental or not, whose position can be determined in measurement. Interactions between fundamental particles will be treated in *Discrete Quantum Mechanics II, Discrete Quantum Electrodynamics*.

Let $T \subset \mathbb{N}$ be a finite (and discrete) time interval such that any particle under study certainly remains in N for $x_0 \in T$. This can always be achieved since, although it may not be possible to prevent particles from leaving N, data from such an event is incomplete, there being no measured final state, and would most reasonably rejected by an experimenter. Without loss of generality let $T = [0, T)$. In each instant a particle does not interact so the state remains the same and is multiplied by a phase, $e^{-iE}$, where $E \in \mathbb{R}$ to preserve the norm. Thus evolution at time *t* is described by the map

$$\mathbb{H}(t) \to \mathbb{H}(t+1) \text{ such that } |f\rangle_{t+1} = e^{-iE}|f\rangle_t \qquad 4.11$$

4.11 is a geometric progression with solution

$$|f\rangle_t = e^{-iEt}|f\rangle_0 \qquad 4.12$$

From 4.6 the general solution of 4.11 at time $t$ is

$$\langle x, t | f \rangle = \left(\frac{1}{2\pi}\right)^{\frac{3}{2}} \int_M d^3p \langle p | f \rangle \, e^{-i(Et - x \cdot p)} \qquad 4.13$$

Phase, $e^{iE}$, is arbitrary and we fix it by defining:

**Definition:** $E = p^0$ is the time like component of a vector $p = (p^0, p)$ at the origin O. $p^0$ is called energy. By definition a mass shell condition $m^2 = p^2 \equiv p \cdot p$ obtains.

**Vector Momentum**

It will be shown in *Discrete Quantum Mechanics II* that 3-momentum is conserved in interactions. Unlike measurement of position this makes it is possible to measure momentum without altering it, so momentum is a vector in the classical sense. $p$ has been defined as a four vector at the origin of an observer's reference frame. A physical vector quantity must be the same for all observers and we define physical energy-momentum as a vector field by parallel displacement (3.18) from the origin

$$p_\mu(x) = p_\gamma(0) k^\gamma{}_\mu(x) \qquad 4.14$$

This is both consistent and unambiguous, because parallel displacement is independent of both coordinate system and path (it is necessary that the effect of displacement from observer A to observer C is the same as that from A to B to C). In particular C may be at an origin in A's future, so quantum covariance effectively determines time evolution, as well as conservation of energy and momentum (Newton's first law) in locally Minkowski regions of space (this is related to Noether's theorem, and like it depends on homogeneity).

**The Wave Function**

Although $\langle x, t | f \rangle$ is discrete in $x$ and $t$, on a macroscopic scale it appears continuous, and it is possible in principle to interpolate between the points of any discrete coordinate system by introducing a new coordinate system (for example by rotating the first). 4.13 can be embedded into a continuous function $f: \mathbb{R}^4 \to \mathbb{C}$, called the wave function defined by

$$\forall x \in \mathbb{R}^4 \quad f(x) = \left(\frac{1}{2\pi}\right)^{\frac{3}{2}} \int_M d^3p \, \langle p | f \rangle e^{-ix \cdot p} \qquad 4.15$$

where the dot product is evaluated at the origin with the Minkowski metric of observer space-time, $\eta = g(0)$, and $p = p(0)$ is a vector. Then we can recover 4.13 for any discrete coordinate system by restricting to N:

$$\forall x \in N, \forall t \in T \quad \langle x | f \rangle = \langle t, x | f \rangle = f(t, x) = f(x) \qquad 4.16$$

Wave functions are not restricted to $\mathcal{L}^2$, but in any reasonable definition of an integral, 2.2 is approximated by the hermitian product in $\mathcal{L}^2$ whenever $f$ and $g$ are in $\mathcal{L}^2$ and the lattice spacing is small.

**The Quantum Derivative**

Classically differentiation means taking values of a function for a small displacement of a parameter. In the quantum context a physically meaningful derivative only makes sense when the function and displacement correspond to measured values, which means taking two measurements infinitesimally close together. But the wave function 4.15 is generally used to relate initial and final measurements which need not be close in time. The covariant derivative is thus not physically applicable, and a covariant wave equation is not a requirement of the model. Analysis of 4.15 reveals

$$P^\mu(0) = -i\partial^\gamma \qquad 4.17$$

so by 4.14 the vector momentum observable is given by

$$P_\mu(x) = -ik^\gamma{}_\mu(x)\partial_\gamma \qquad 4.18$$

This is the derivative under quantum covariance, found by formally differentiating the wave function and identifying the result for different observers using parallel displacement. The quantum derivative 4.18 acting on an eigenstate of momentum uniquely defines a true vector field, whereas, as is well reported in the literature, the general covariant derivative leads to a path dependency in momentum with consequent ambiguities to conservation laws, and it introduces curvature terms which create difficulties for second quantisation. This does not occur for the quantum derivative used here.

It is immediately possible to write down the Dirac equation and Klein-Gordon equation using the quantum derivative. The solution will not be identical to that found with a conventional covariant derivative, but we can only expect an experimental difference in experiments which exhibit specifically quantum behaviour in strong gravitational fields.

**Bounds**

Units have been used such that the components of momenta are bounded by $\pm\pi$. The value of this bound in conventional units depends on lattice spacing according to the formula

$$p_{\max} = 2\pi\hbar/\text{spacing}$$

While it might be possible in principle to carry out measurement of momentum with large lattice spacing and carry out an error analysis for quantum mechanics using the DFT, the theoretical bound on momentum depends on the limit of small lattice spacing, not on the actual lattice used in a given instance. Thus in practice we are more interested in small lattice spacing, and whether there is some non-zero lower bound. The remaining sections of this paper indicate the existence a fundamental discrete unit of proper time, $\chi$, between particle interactions of magnitude twice the Schwarzschild radius $4Gm/c^3$ for an elementary particle of mass $m$ where $G$ is the gravitation constant. For an electron

$$\chi = 4Gm/c^3 = 9.02 \times 10^{-66} \text{ sec} \qquad 4.19$$

This leads to a bound of $2.29\times10^{52}$ eV or $4.075\times10^{14}$ kg for the energy of a single electron, well beyond any reasonable energy level.

It makes little difference whether discreteness is taken into account in most practical measurement. For example, the components of the discrete momentum operator are given by

$$\text{for } i = 1, 2, 3, \ \bar{P}^i = \frac{-i}{2}\sum_{x \in N}|x\rangle[\langle x+1^i| - \langle x-1^i|]$$

Then

$$P^i|p\rangle = \frac{-i}{2}\sum_{x \in \mathbb{N}}|x\rangle[\langle x+1^i| - \langle x-1^i|]|p\rangle = \sum_{x \in \mathbb{N}}|x\rangle\langle x|p\rangle\sin p^i = \sin p^i|p\rangle \qquad 4.20$$

So the eigenvalue of momentum is $\sin p \approx p$ for $p$ much less than the bound of $\pi/\chi$. Then an electron with a difference of 0.1% between $p$ and $\sin p$ would have an energy of $0.055\pi/\chi = 1.38 \times 10^{52}$ eV, which seems unrealistic.

**Manifest Covariance**

Because momentum space is bounded by M the local flat space approximation of 4.15 is not manifestly covariant, but under reasonable conditions it is covariant for physically realisable states and transformations. To see this we observe that conservation of energy follows directly from parallel displacement of energy momentum. It follows that the total energy of a system is bounded provided only that energy has been bounded at some time in the past. This is the case whenever an energy value is known since a measurement of energy creates an eigenstate with a definite value of energy. Momentum is bounded by energy and the mass shell condition. The probability of finding a momentum above this bound is zero, and we assume that for physically realisable states the discrete representation of $\langle\overline{p}|f\rangle$ has support which is bounded in each component of momentum. The bound depends on the system under consideration, but without wishing to specify it, we observe that, provided that the discrete unit of time is sufficiently small, momentum is always much less than $\pi/4$.

Physically meaningful Lorentz transformation cannot boost momentum beyond the absolute bound of $\pi$ because physical reference frames are defined with respect to macroscopic matter. A realistic Lorentz transformation means that macroscopic matter has been physically boosted by the amount of the transformation. For example for a cubic lattice with spacing equal to the Schwarzschild radius of an electron a boost in the order of $\pi/4$ would require an energy of $2 \times 10^{14}$ solar masses per kilogram of matter to be boosted. Thus we can ensure manifest covariance by imposing the condition that all momentum space wave functions have a representation in a subset of M bounded by some realistic energy level. Then we remove the non-physical periodic property of $\langle p|f\rangle$ by replacing

$$\Theta_M(p)\langle p|f\rangle \rightarrow \langle p|f\rangle$$

where $\Theta_M(p) = 1$ if $p \in M$ and $\Theta_M(p) = 0$ otherwise. 4.15 is then replaced with the standard form, up to normalisation of $\langle p|f\rangle$, of the wave function in relativistic quantum mechanics

$$f(x) = \left(\frac{1}{2\pi}\right)^{\frac{3}{2}}\int_{\mathbb{R}^3} d^3p\, \langle p|f\rangle\, e^{-ix\cdot p} \qquad 4.21$$

**Lorentz Transformation**

Let $|x\rangle$ be a state of a particle at definite position $x$ in the lattice at some time $x^0$. Then, from 4.9, the Lorentz transformation is

$$|\Lambda x\rangle = \int_{\mathbb{R}^3} d^3p|p\rangle\langle p|\Lambda x\rangle = \int_{\mathbb{R}^3} d^3p|p\rangle\langle \Lambda' p|x\rangle$$

So we have for $\Lambda$

$$\int_{\mathbb{R}^3} d^3p|\Lambda p\rangle\langle p| = \Lambda = \int_{\mathbb{R}^3} d^3p|p\rangle\langle \Lambda' p| \qquad 4.22$$

We impose a primed coordinate system at time $t' = x^{0'}$ after transformation by restricting the wave function to points $x'$ in a new cubic lattice N'. Then the transformed state is the restriction to the new lattice, i.e

$$|\Lambda x\rangle = \sum_{x' \in N'} |x'\rangle\langle x'|\Lambda x\rangle \qquad 4.23$$

$|\Lambda x\rangle$ is not an eigenstate of position in N'; if a measurement of position were done in N' and we were then to transform back to N the state would no longer be $|x\rangle$. I.e. the operators for position in different frames N and N' do not commute. But if no measurement is done, it is straightforward to show that we can transform straight back and recover $|x\rangle$,

$$\Lambda'|\Lambda y\rangle = |y\rangle \qquad 4.24$$

so there is no problem with lack of unitarity under Lorentz transformation.

**Geodesic Motion**

In the classical correspondence the motion may be described as a sequence of states at instances separated by some time interval $\chi$ which is sufficiently small that there is negligible alteration in predictions in the limit in which $\chi$ tends to zero. The state at each instant may be regarded as the initial state, and state at the next instant may be regarded as the final state, which becomes the initial state for the next part of the motion. The motion between instances is described by a wave function and momentum is infinitesimally parallel displaced from the initial to the final state. The cumulative effect of infinitesimal parallel displacements is parallel transport. For a classical body or for a ray of light during each small time interval the wave function is confined in both momentum space and coordinate space (to the required accuracy) and since intermediate states are determinate and may be calculated in principal the intermediate states are effectively measured and may be regarded as measured states for the interpretation of quantum theory. Then for a classical body or for a ray of light the motion in spacetime is in the direction of $p$, unaffected by observation. Geodesic motion follows as the cumulative effect of parallel displacement over small time increments.

It should be no surprise that geodesic motion follows in the classical correspondence for non-interacting particles, since the same is true for general relativity as shown by Einstein et al. (reported in [6]). However that derivation depends on the Bianchi identities which do not apply here because the covariant derivative breaks down in the quantum domain.

## 5   Gravity

**The Principle of Equivalence**

In Newtonian mechanics we may distinguish between force, impressed force and inertial force. A force is anything which causes an acceleration according to the second law. Impressed force implies the physical action of one body on another, including action by means of the physical transmission of an intermediary. An *inertial* (from latin *not acting*) force is caused not by the action of one body on another but by the choice of reference frame. The coriolis and centrifugal forces in rotating frames, and *g*-forces in accelerated frames are inertial. Inertial forces are distinguished from impressed forces since they are not an action and produce no third law reaction. Because the coriolis and centrifugal forces do not exist in inertial reference frames they are often regarded as fictional, but a general treat-ment of reference frames incorporates both inertial and non-inertial frames, and inertial forces may be said to exist in non-inertial frames by the second law definition of force. In curved space-time Newton's first law is restored by the definition *gravity is the inertial force appearing in an inertial observer space-time*. For example the acceleration of inertial matter due to gravity at the surface of the earth will be understood in an inertial observer space-time with an origin at the centre of the earth. Then local reference frames fixed at the surface of the earth accelerate relative to matter in free fall, and hence *gravity is equivalent to the g-force caused by the acceleration of a local reference frame relative to inertial matter*. This is the principle of equivalence.

**Gravitational Red Shift**

According to the principle of general relativity the laws of physics in any locality are the same as the laws of physics in any other locality. In each locality the laws of physics are expressed with respect to time determined by an unperturbed or inertial clock in that locality. We have no evidence on which to assume that clocks at different places can be synchronised even if they are stationary with respect to each other. A second is a second as defined by the behaviour of $Cs^{133}$ in a particular location. It cannot be assumed that this will be synchronised with a second determined by $Cs^{133}$ in a different place. As with the special theory the relationship between the rate at which the clocks determine the time is measured by red shift. Special relativity studied the consequences of red shift due to difference in speed of clocks, whereas general relativity treats the effects due to difference in location.

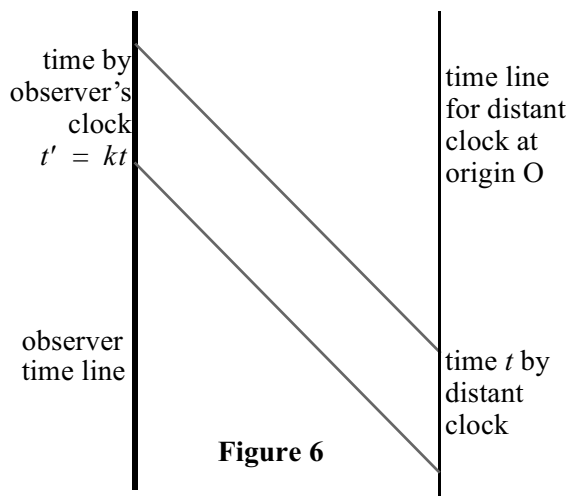

**Figure 6**

Consider first a static system. Moving clocks may be treated by observing that space local to the clock is Minkowski and applying Lorentz transformation. A space-time diagram (figure 6), is drawn using observer space-time, such that light is drawn at 45° and lines of equal time are horizontal. The observer measures primed locally Minkowski coordinates $x^{\mu'}$ referred to his clock, and locally Minkowski unprimed coordinates $x^\mu$ are set up referring to the distant clock at an origin O. As with the Doppler shift used in the *k*-calculus for special relativity, the time interval between two signals sent from the distant clock to the observer is multiplied by red shift, $k$, $x^{0'} = kx^0$ (by considering the signals as the start and stop of a burst of light of a set number of wavelengths of a set frequency). For red shift $k > 1$ the observer's clock is faster, so that, in the observer space-time of the distant clock, the coordinate distance between the ends of a rod near the observer is proportionately shorter that the metric distance, or real length of the rod measured locally by the observer from the definition based on light speed. Hence $x^1 = kx^{1'}$. Due to spherical symmetry there is no change in the relationship between metric distance and coordinate distance associated with rotation, so the general transformation law $x^{\mu'} = x^{\mu'}_{,\nu} x^\nu$ leads to

$$k^\mu{}_\nu = \begin{bmatrix} k & 0 & 0 & 0 \\ 0 & 1/k & 0 & 0 \\ 0 & 0 & 1 & 0 \\ 0 & 0 & 0 & 1 \end{bmatrix} \qquad 5.1$$

So the coefficients $k^\mu{}_\nu$ depend entirely on a single measurable quantity, namely red shift. More generally $k^\mu{}_\nu$ for a moving observers is found from the static case by applying Lorentz transformation and the net red shift contains both gravitational and Doppler parts.

**A Correction to Radar**

A natural modification to the analysis of radar is to hypothesise a time delay between absorption and emission in proper time of a fundamental particle (typically an electron) reflecting electromagnetic radiation. Indeed it has often been suggested that such a small scale correction could justify the cut-off in the treatment of loop divergences and resolve the problem of the Landau pole, and if particle interactions are discrete it is reasonable to anticipate a small delay. This is not the usual assumption of general relativity, but it will be shown that it leads to Schwarzschild geometry and hence to Einstein's field equation. In fact, if the reflection of a photon were really instantaneous photon exchange would give a fixed metric, which may be Minkowski or it may be a metric with constant curvature. Clearly this is not true for our universe, so it seems reasonable to look at ways of modifying the analysis.

To affect the geometry of space-time the delay between absorption and emission must be a fundamental property of nature which must be taken into account in addition to speed in determining the shortest time for the transmission of information. The metric is still determined as in special relativity by the minimum time for the return of information reflected at an event, but now this minimum net time depends not only on the maximum theoretical speed of information but also on the minimum possible time between absorption and emission in the reflection of a photon. This minimum time is fundamental property of the interactions between particles, and will be calculated from the gravity due to a single elementary particle (electron or quark). It is interesting to observe also that the hypothesis of proper time delay between interactions also relates gravitational mass to inertial mass. If, for example, the interactions of a muon are identical to, but much less frequent than, the interactions of an electron then we would expect to observe a proportionately smaller change in motion from the same stimulus. We postulate a proper time delay $4GM$, where $4G$ is a constant of proportionality. Later $G$ will be identified with the gravitational constant.

In *Discrete Quantum Mechanics II*, and in contrast to the instantaneous reflection of a classical e.m. wave, the reflection of a photon is treated as two events, absorption and emission. Ordinarily in QED, emission can occur at any time, before or after absorption, and it is necessary to integrate over all such possibilities. In the particular case under study there is a measurement and hence an eigenstate of position. In the classical limit the integration gives a superposition of states equivalent to the instantaneous reflection of a classical e.m. wave. In terms of the path integral, other paths contribute nothing to the amplitude and on the remaining path the photon is instantaneously reflected. But according to the interpretation given above, in an eigenstate of position the photon follows a definite, known path describable by space-time diagrams, and in this eigenstate the time of reflection is small but not instantaneous.

**A Curved Space-time Diagram** (figure 7)

A single elementary particle at an exact position (i.e. in an eigenstate) at an origin O has spherical symmetry and space-time diagrams may be used to show a radial coordinate in *n* dimensions without loss of generality. The reflection of a radar pulse sent by a distant observer is now seen as two events, absorption, A, and emission, E. The reflection appears instantaneous in a frame using primed coordinates defined by an observer using radar, but in proper time and in observer space-time at O with unprimed coordinates the time between them is $4GM$, as shown in the diagram. According to the definition of distance by radar the position coordinate of the observer in unprimed coordinates defined from a clock at O is

$$r = (t_2 - t_1)/2 \qquad 5.2$$

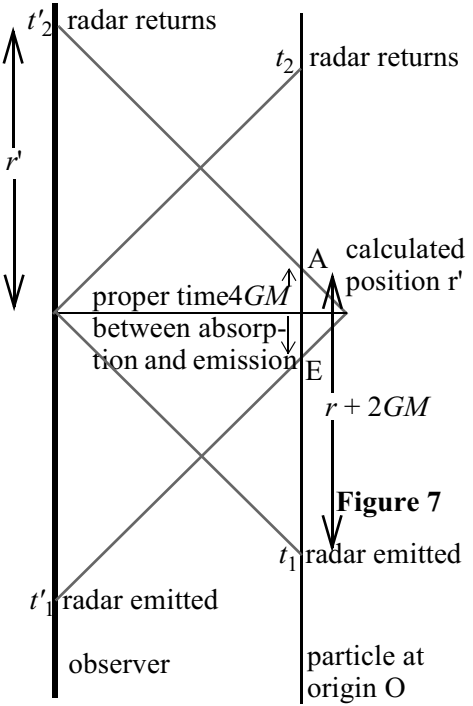

Figure 7

But in primed Minkowski coordinates using the observer's clock the position coordinate of O is

$$r' = (t'_2 - t'_1)/2 \qquad 5.3$$

and using red shift as defined in figure 6

$$r' = k(r + 2GM) \qquad 5.4$$

**Schwarzschild**

We seek to analyse the gravitational effect of a single elementary particle at O in figure 8. Place the gravitating particle at the origin O in observer space-time at O with unprimed radial coordinates. The observer uses primed radial coordinates with the origin translated to O, so that $k^\mu{}_\nu$ is given by 5.1. Draw a synchronous slice through the gravitating particle in observer space-time at O and superpose the primed coordinates (figure 8). The observer's radial coordinate is stretched so that the origins coincide and the particle is shown as a point in the diagram. Consider an infinitesimal coordinate length $ds$ on the circumference. Using 5.4

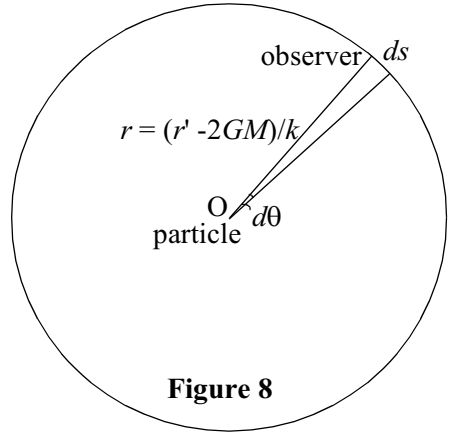

Figure 8

$$ds = rd\theta = (r' - 2GM)d\theta/k \qquad 5.5$$

For $k > 1$ (red shift) the observer's clock runs fast, so distances measured by proper time along the circumference are correspondingly shorter. Hence in the primed coordinates (since $\theta = \theta'$)

$$ds = kr'd\theta \qquad 5.6$$

Comparing 5.6 and 5.5

$$k^2 = 1 - \frac{2GM}{r'} \qquad 5.7$$

From 3.12, 5.1 and 5.7 we find the familiar form of the Schwarzschild metric,

$$g_{\alpha\beta} = \eta_{\mu\nu} k^{\mu}{}_{\alpha} k^{\nu}{}_{\beta} = \begin{bmatrix} 1 - \frac{2GM}{r'} & 0 & 0 & 0 \\ 0 & -\left(1 - \frac{2GM}{r'}\right)^{-1} & 0 & 0 \\ 0 & 0 & -r'^2 & 0 \\ 0 & 0 & 0 & -r'^2 \sin^2\theta \end{bmatrix} \qquad 5.8$$

**Black Holes and Singularities**

This view of geometry puts a different light on the singularity at the centre of Schwarzschild because in information space curvature is defined only with respect to observer space-time. The event horizon marks a bound at which this relationship becomes singular and beyond which observer space-time is not defined and geometry ceases to be physically meaningful; information space does not exist beyond the event horizon. This view does not deny the mathematical possibility of a mathematical extension to the manifold inside the event horizon, but it does deny that any such extension has a physical meaning. Figure 7 shows the interactions of an elementary particle in an eigenstate of position taking place on the event horizon for that particle. When many particles have the same position, as in a black hole, the geometrical effect is to increase the radius of the event horizon at which all interactions take place. This model might be expected to modify the calculation of Hawking radiation. Qualitatively the process is the same, but the negative energy particle is not required to fall through the event horizon, but interacts at it. I have not done a calculation, and in the absence of experimental data such a difference from semi-classical general relativity appears to be academic.

**The Newtonian Approximation**

Vector momentum is given by 4.14 so that the corresponding coordinate space vector (3.7) is constant

$$k^{\alpha}{}_{\beta}(x) p^{\beta}(x) = \text{const} \qquad 5.9$$

With 5.1 this affects the frequency of the wave function, and hence the energy of a particle, according to the relation

$$k p^0 = \text{const} \qquad 5.10$$

So the classical energy $E$ satisfies

$$E = \langle P^0 \rangle \propto 1/k = \left(1 - \frac{2GM}{r'}\right)^{-1/2} \approx 1 + \frac{GM}{r'} \qquad 5.11$$

For a body of mass $m$ the constant of proportionality is fixed at $r' = \infty$

$$E = m + \frac{GMm}{r'} \qquad 5.12$$

showing the gravitational potential in the Newtonian approximation

**Measurement of Curvature**

Eppley and Hannah [12] argued that if a gravitational measurement causes wave function collapse, violation of the uncertainty relationships can only be avoided if by giving up conservation of momentum, whereas if a gravitational measurement does not cause wave function collapse, then they can be used to send observable signals faster than light (by measuring components of entangled wave functions; the usual argument that this cannot be used to transmit information fails if one can perform a measurement without wave function collapse). Page & Geilker [19] set up a Cavendish experiment with the locations of the test spheres dependent upon the outcome of a quantum process. They found that the gravitational effects always come from actual locations of the masses not the average as would be expected for a mixed state, but this.casts little light on the issue, since it is not possible to say clearly that the macroscopic masses are correctly described by a mixed state rather than by a classical probability governed by the result of a quantum process, where the result is treated as a random variable.

The interpretation given in this paper is that classical bodies are always governed by classical laws derived statistically from the laws of quantum mechanics applied to systems of many particles, so that one would not achieve a true mixed state in the Page-Geilker experiment until a sufficient number of years have passed for statistical uncertainties to become significant, and then only if the massive gravitating bodies were truly isolated. In the present account collapse is simply a change in information, and any measurement whose result is not determinate provides information and causes collapse. There is an argument that a determination of gravity of sufficient accuracy to determine the position of a gravitating particle is effectively a measurement of position and causes a collapse of that particle's wave function. However we do not apply Eppley & Hannah's argument because the momentum field is determined from parallel displacement, not from a covariant differential equation. When the wave function of the gravitating body collapses the information we have about curvature alters, and the manifold is altered, but parallel displacement again determines momentum unambiguously from 4.14. The change in the manifold under collapse is not a problem for the theory because it is merely a change in information space, not to ontology.

We ignore further discussion of the effect of measurement of curvature and consider instead the curvature of the manifold between measurements. The Einstein tensor $G^{\alpha\beta}$ determines $k^{\mu}_{\nu}$ as it does in the classical theory, and hence the evolution of the wave function between measurements which is ultimately responsible for the effect of gravity. $G^{\alpha\beta}$ will be determined by the information about mass distribution available at time of the initial measurement. Information from a subsequent measurement modifies $G^{\alpha\beta}$ as appropriate to the next part of the motion.

**Einstein's Field Equation**

In the static system shown in figure 6 there may be two causes of gravitational red shift; $k$ may be a function of position or distance in empty space, and $k$ may be directly dependent on the distribution of matter. In either case a tensor equation is required to describe geometry, and it is convenient to use an equation for curvature. In the vacuum case $k$ is a function of distance, and from the homogeneity of the vacuum it follows that this is described by constant curvature, which gives the cosmological constant term in Einstein's Field equation for empty space,

$$R_{\mu\nu} = \lambda g_{\mu\nu} \qquad 5.13$$

Ignoring the cosmological constant it is now possible to follow the standard argument that the Einstein tensor is proportional to stress energy

$$G^{\alpha\beta} = R^{\alpha\beta} - \tfrac{1}{2}g^{\alpha\beta}R = 8\pi G \langle T^{\alpha\beta} \rangle \qquad 5.14$$

but a purpose of this paper is to see that Einstein's field equation 5.14 follows from the definition of a metric in terms of two way light speed, after taking into account a small delay in reflection. It has been shown that Schwarzschild geometry obtains for a single point particle in an eigenstate of position. Schwarzschild is a known solution of Einstein's field equation so we have

$$G^{\alpha\beta}|x\rangle = 8\pi G T^{\alpha\beta}|x\rangle \qquad 5.15$$

By linearity

$$G^{\alpha\beta}|f\rangle = 8\pi G T^{\alpha\beta}|f\rangle \qquad 5.16$$

and the result extends immediately to states of $n$ particles in the appropriate Fock space. Thus generally

$$G^{\alpha\beta} = 8\pi G \langle T^{\alpha\beta} \rangle \qquad 5.17$$

The energy density operator is given by 4.18

$$P_0(x) = -ik^{\gamma}{}_0(x)\partial_{\gamma} \qquad 5.18$$

and the general form of the stress energy tensor is found by composing a second rank tensor which reduces to energy in the rest frame of a particle. This can be done in a straightforward way for electrons and photons after constructing the corresponding field operators, as in *Discrete Quantum Mechanics II*.

## Acknowledgements

I should like to thank a number of physicists who have (often unwittingly) contributed to the development, content and presentation of this paper, particularly Marcus Appleby, John Farina, Steve Carlip, Eric Forgy, Paul Colby, Matthew Nobes, Mike Mowbray, Frank Wappler, Jim Carr, Chris Hillman and John Baez and the moderators of sci.physics.research (John Baez, Matt McIrvin, Ted Bunn & Kevin Scaldeferri) for their vigilence in pointing out lack of clarity in expression, as well as for maintaining a public forum of sufficient expertise to provide valuable feedback on the presentation and properties of the model